\documentclass{iaus}
\usepackage{graphicx}

\title[Mass and orbit constraints of LS~5039] 
{Mass and orbit constraints of the gamma-ray binary LS~5039}

\author[Szalai et al.]   
{T. Szalai$^1$, G. E. Sarty$^{2,3}$, L. L. Kiss$^4$, J. M. Matthews$^5$, J. Vink\'o$^1$, \\and Cs. Kiss$^4$} 
\affiliation{$^1$Department of Optics and Quantum Electronics, University of Szeged, \\ D\'om t\'er 9., Szeged H-6720, Hungary;
email: {\tt szaszi@titan.physx.u-szeged.hu} \\[\affilskip]
$^2$Royal Astronomical Society of Canada, Saskatoon Centre, 
\\P.O. Box 317, RPO University, Saskatoon, SK S7N 4J8, Canada \\[\affilskip]
$^3$Department of Physics and Engineering Physics, University of Saskatchewan, \\Saskatoon, SK S7N 5E2, Canada \\[\affilskip]
$^4$Konkoly Observatory of the Hungarian Academy of Sciences, \\H-1525 Budapest, P.O. Box 67, Hungary \\[\affilskip]
$^5$Department of Physics and Astronomy, University of British Columbia, 
\\6224 Agricultural Road, Vancouver, BC V6T 1Z1, Canada
}
\begin{document}
\maketitle
\begin{abstract}
We present the results of space-based photometric and ground-based spectroscopic observing campaigns on the
$\gamma$-ray binary LS~5039. The new orbital and physical parameters of the system are similar to former results, except
we found a lower eccentricity.
Our {\em MOST}-data show that any broad-band optical photometric variability at the orbital period is below the 2 mmag
level. Light curve simulations support the lower value of eccentricity and imply that the mass of the compact object is higher than 
1.8 M$_{\odot}$.

\end{abstract}

\section{Introduction}
LS~5039, the enigmatic high-mass X-ray binary has been intensively observed 
at various wavelengths in the past years (see \cite[Sarty et al. 2011, hereafter S11]{Sarty_etal11} for a review). 
\cite[Paredes et al. (2000)]{Paredes_etal00} identified relativistic radio jets and
also a very high energy (VHE) gamma-ray source at the coordinates of the system; therefore LS~5039 became one of a handful of known gamma-ray binaries.
There are several open issues about the system (see S11), but the major question is 
whether the secondary component orbiting around the O6.5V((f)) star is a black hole or a non-accreting young pulsar.

Hereinafter we show the results of our spectroscopic and photometric analysis concerning 
mass and orbit constraints of LS~5039 (see the details in S11).

\section{Analysis and parameter determination}
Spectroscopic observations were carried out in 2009 with the echelle spectrograph mounted at ANU 2.3m Telescope 
(SSO, Australia), and in 2011 using FEROS (\cite[Kaufer et al. 1999]{Kaufer_etal99}) at MPG/ESO-2.2m telescope at La Silla, Chile. 
Covering $\sim$40 hours with nearly uniform sampling of the whole orbit between 3900-6750 \AA\ with a resolving power 
$\lambda$/$\Delta \lambda \approx$ 23,000 at H$\alpha$, it is the highest resolution, homogeneous spectral dataset ever obtained 
for LS~5039.

Radial velocities (RV) of H\begin{small}I\end{small} and He\begin{small}I\end{small} lines show a systematic blueshift 
with respect to the RVs of He\begin{small}II\end{small} lines, therefore only the latter ones were used to fit eclipsing binary models 
with the Wilson--Devinney (WD) code (\cite[Wilson \& van Hamme 2003]{Wilson_vanHamme_03}).
We do not see signs of non-radial pulsations in our data in contrast to the results reported by \cite[Casares et al. (2010)]{Casares_etal10},
see Fig. 1.

\begin{figure}
\begin{center}
\includegraphics[height=4cm]{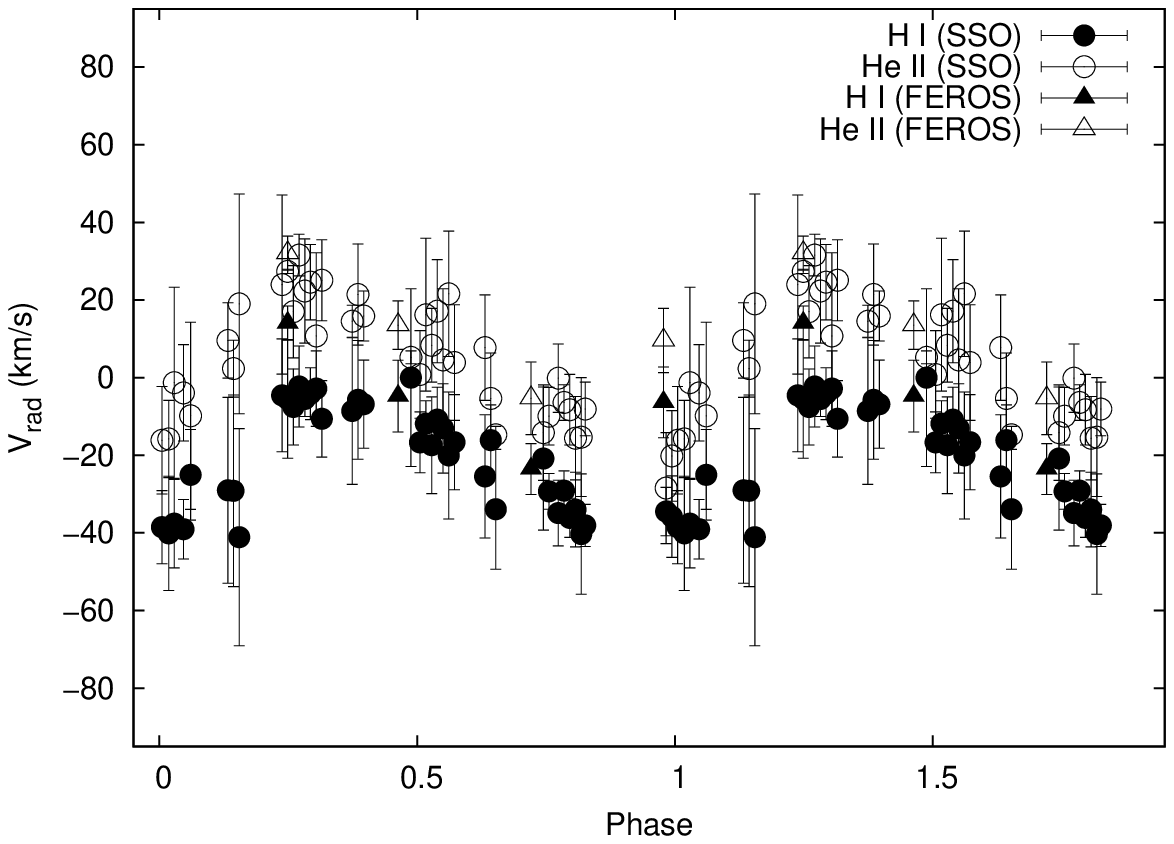} \hskip 2mm
\includegraphics[height=4cm]{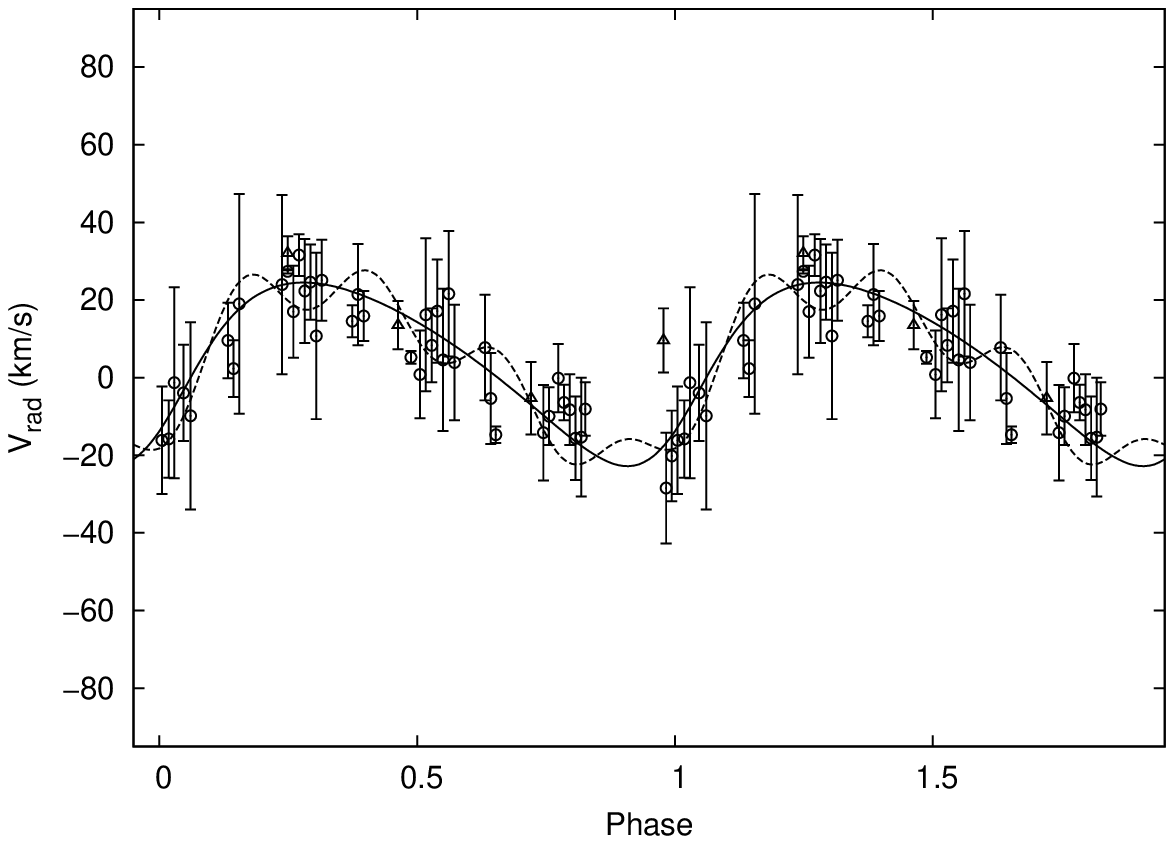}
\caption{{\it Left}: Radial velocities based on H Balmer and HeII lines. {\it Right}: The best-fitting curve to radial velocities of
HeII lines (solid line), also with the assumed pulsation of the O star (dotted line).}
\label{fig:1}
\end{center}
\end{figure}

Our data were analyzed as described in S11 but now with the addition of the FEROS data. 
Orbital parameters are close to previous solutions (\cite[Casares et al. 2005]{Casares_etal05}, \cite[Aragona et al. 2009]{Aragona_etal09}), 
but we found the orbital eccentricity ($e$=0.24 $\pm$ 0.08) being definitely lower than determined previously.

Photometric data, obtained with {\em MOST} satellite in July of 2009, indicates a variability at the level of 2 mmag.
Our light curve (LC) simulations show that the amplitude of LCs decreases with increasing total mass or 
decreasing eccentricity. Comparison of data and models suggests that primary mass is at the higher end of estimates based on its 
spectral type ($\sim$26 M$_{\odot}$), from which we get a mass for the compact star of at least 1.8 M$_{\odot}$.

\section{Conclusions}
We carried out a detailed spectroscopic and photometric analysis to get mass and orbit constraints of the $\gamma$-ray binary LS~5039.
The new system parameters are close to previously determined ones, except that we found a lower eccentricity. LC simulations support this result, and
imply that the compact object may be a low mass black hole -- but do not fully exclude that it may be a neutron star.

\section*{Acknowledgments} 
This work has been supported by the Australian Research Council, the University
of Sydney, the Hungarian OTKA Grant K76816 and the ``Lend\"ulet'' Young
Researchers' Program of the Hungarian Academy of Sciences.

\end{document}